\documentclass[journal=jacsat,manuscript=article]{achemso}
\usepackage{textcomp,gensymb,amsmath,xcolor,physics,pdfpages}
\newcommand{\bc}{\begin{center}}
	\newcommand{\ec}{\end{center}}
\newcommand{\vs}{\vspace}

\newcommand{\beq}{\begin{equation}}
\newcommand{\eeq}{\end{equation}}
\newcommand{\beqs}{\begin{eqn*}}
	\newcommand{\eeqs}{\end{eqn*}}
\newcommand{\bq}{\begin{quote}}
	\newcommand{\eq}{\end{quote}}
\newcommand{\beqa}{\begin{eqnarray}}
\newcommand{\eeqa}{\end{eqnarray}}
\newcommand{\beqas}{\begin{eqnarray*}}
	\newcommand{\eeqas}{\end{eqnarray*}}
\newcommand{\bfg}{\begin{figure}}
	\newcommand{\efg}{\end{figure}}

\newcommand{\tu}{\textsuperscript}
\newcommand{\tb}{\textsubscript}
\newcommand{\txc}{} 
\newcommand{\txb}{} 

\author{Sarthak Das}
\affiliation[IIScECE]
{Department of Electrical Communication Engineering, Indian Institute of Science, Bangalore 560012, India}
\author{Medha Dandu}
\affiliation[IIScECE]
{Department of Electrical Communication Engineering, Indian Institute of Science, Bangalore 560012, India}
\author{Garima Gupta}
\affiliation[IIScECE]
{Department of Electrical Communication Engineering, Indian Institute of Science, Bangalore 560012, India}
\author{Krishna Murali}
\affiliation[IIScECE]
{Department of Electrical Communication Engineering, Indian Institute of Science, Bangalore 560012, India}
\author{Nithin Abraham}
\affiliation[IIScECE]
{Department of Electrical Communication Engineering, Indian Institute of Science, Bangalore 560012, India}
\author{Sangeeth Kallatt}
\affiliation[NBI]
{Center for Quantum Devices, Niels Bohr Institute, University of Copenhagen, Denmark}
\author{Kenji Watanabe}
\affiliation[NIMS1]
{Research Center for Functional Materials, National Institute for Materials Science,
1-1 Namiki, Tsukuba 305-044, Japan}
\author{Takashi Taniguchi}
\affiliation[NIMS2]
{International Center for Materials Nanoarchitectonics, National Institute for Materials Science,
1-1 Namiki, Tsukuba, 305-044 Japan}
\author{Kausik Majumdar}
\email{kausikm@iisc.ac.in}
\affiliation[IIScECE]
{Department of Electrical Communication Engineering, Indian Institute of Science, Bangalore 560012, India}
\title{Highly tunable layered exciton in bilayer WS$_2$: linear quantum confined Stark effect \emph{versus} electrostatic doping}

\begin{document}

{\abstract In 1H monolayer transition metal dichalcogenide, the inversion symmetry is broken, while the reflection symmetry is maintained. On the contrary, in the bilayer, the inversion symmetry is restored, but the reflection symmetry is broken. As a consequence of these contrasting symmetries, here we show that bilayer WS\tb2 exhibits a quantum confined Stark effect (QCSE) that is linear with the applied out-of-plane electric field, in contrary to a quadratic one for a monolayer. The interplay between the unique layer degree of freedom in the bilayer and the field driven \txb{partial} inter-conversion between intra-layer and inter-layer excitons generates a giant tunability of the exciton oscillator strength. This makes bilayer WS\tb2 a promising candidate for an atomically thin, tunable electro-absorption modulator at the exciton resonance, particularly when stacked on top of a graphene layer that provides an ultra-fast non-radiative relaxation channel. By tweaking the biasing configuration, we further show that the excitonic response can be largely tuned through electrostatic doping, by efficiently transferring the oscillator strength from neutral to charged exciton. The findings are prospective towards highly tunable, atomically thin, compact and light, on chip, reconfigurable components for next generation optoelectronics.}
\newpage
The ultra-fast radiative recombination rate \cite{robert2016exciton,palummo2015exciton,gupta2019fundamental} of the strongly bound excitons \cite{he2014tightly,chernikov2014exciton,gupta2017direct} in monolayer transition metal dichalcogenides (TMDCs) leads to efficient photoemission \cite{splendiani2010emerging,paur2019electroluminescence}. The fast re-radiation of the absorbed photon by the exciton without ample opportunity to suffer from non-radiative processes results in near perfect excitonic reflectance \cite{scuri2018large,back2018realization,zhou2020controlling}. On the other hand, in a bilayer TMDC, a lower indirect energy bandgap appears while the direct energy gap at the $K(K^\prime)$ point of the Brillouin zone is maintained. This causes the \txc{momentum-direct} excitons in bilayer to encounter strong non-radiative processes \cite{das2019layer}, and thus a bilayer TMDC is weakly luminescent \cite{mak2010atomically}, while still exhibiting strong excitonic absorption.

To be able to tune the excitonic absorbtion in a dynamic manner is crucial to multiple applications such as programmable optical modulators, tunable photodetectors, optical computing and imaging, laser pulse shaping, and ultra-light optomechanical mirrors. A bilayer distinguishes itself from the monolayers in a number of ways in terms of tunability of the exciton oscillator strength and its spectral control. First, the reflection symmetry along the c-axis is present in a monolayer TMDC, but is broken in a bilayer. Thus, in presence of a vertical electric field, one expects that a bilayer would show linear Stark shift, in contrast to a quadratic one \cite{verzhbitskiy2019suppressed,roch2018quantum,massicotte2018dissociation,chakraborty2019electrically} in monolayer. In spite of theoretical predictions of linear quantum confined Stark effect (QCSE) in bilayer TMDC \cite{liu2012tuning,lu2017stark,nguyen2016band,ramasubramaniam2011tunable,yang2010modulation,zibouche2014transition,shanavas2015effective,zhang2014indirect}, there has not been any experimental demonstration of the same to date for the direct \txc{intra-layer} $A$ exciton. Second, the presence of two layers in a bilayer, which are rotated at $180^\circ$ with respect to each other, offers an additional layer degree of freedom \cite{das2019layer}. With a vertical electric field, there is a possibility to convert the intra-layer exciton at zero field to an inter-layer exciton, resulting in a strong tunability of the oscillator strength. Third, it is easier to inject carriers in a bilayer from a metal contact due to reduced bandgap, increased valley degeneracy and enhanced orbital coupling to metal \cite{kim2012high}. This allows fast and efficient modulation of the excitonic oscillator strength by inter-conversion between a neutral exciton and a charge exciton (trion) through electrostatic doping \cite{pei2015exciton,kummell2015gate,wu2013electrical}.

In this work we use two different device structures to segregate the effects of electric field induced QCSE and electrostatically induced doping on both the strength and the spectral tunability of the excitonic response in a bilayer WS\tb2. We use a few-layer graphene (FLG)/hBN/2L-WS\tb2/hBN/FLG stack keeping the bilayer WS\tb2 electrically floating to report linear QCSE of the direct \txc{intra-layer} exciton. The corresponding excitonic oscillator strength is shown to exhibit a large tunability due to an electric field driven \txb{partial} inter-conversion between intra-layer and inter-layer exciton. We further demonstrate that placing a graphene layer right under the bilayer WS\tb2 is promising for an atomically thin, gate tunable electro-absorption layer. To demonstrate the electrostatically induced doping effect, we use a similar stack, but with the bilayer WS\tb2 connected to a grounded metal contact. The efficient carrier injection from the contact leads to effective oscillator strength transfer from neutral exciton to positive or negatively charged trions depending on the polarity of the gate voltage.

\begin{figure}[!hbt]
		\centering
		\vs{-0.1in}
		\includegraphics[scale=0.5]{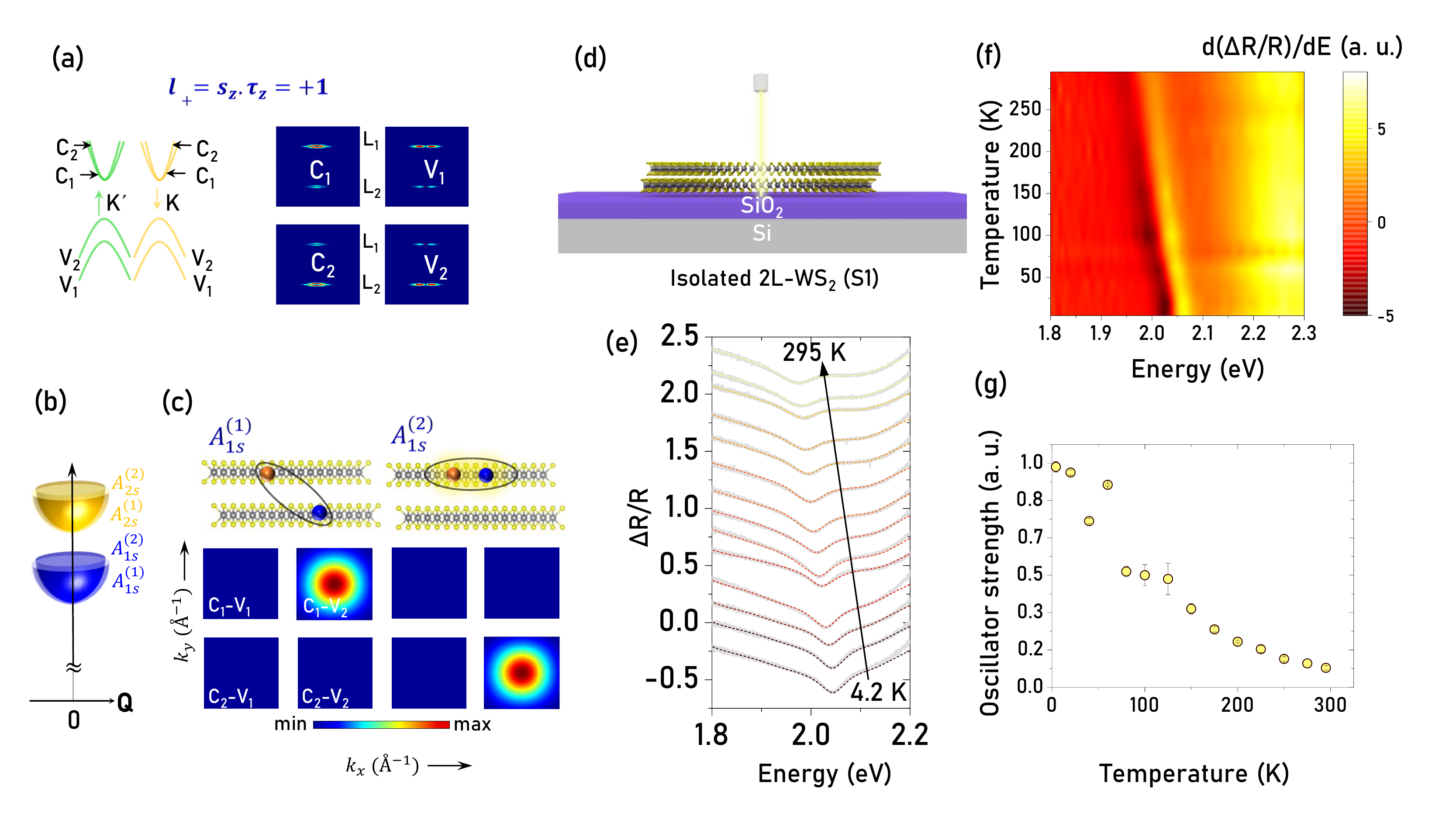}
		\caption{\textbf{Intra- and inter-layer direct exciton in bilayer WS\tb2.}
			(a) Left panel: Lowest energy bands of bilayer WS\textsubscript{2} for a given spin-valley $(l_+=s_z.t_z=+1)$ configuration around $\mathbf{K}$ ($\mathbf{K^\prime}$) point. Right panel: The corresponding layer distribution of the electron and hole wave functions for $\mathbf{k} = \mathbf{K} + \Delta\mathbf{k}$ where $\Delta\mathbf{k}=\frac{2\pi}{a}[0.0033,0.0033]$, $a$ is the lattice constant.
			(b) Exciton energy band structure as a function of center of mass momentum ($\mathbf{Q}$) showing different 1s and 2s bands.
            (c) The $k$ space distribution around $\mathbf{K}$ for both the inter-layer \txc{[$A_{1s}^{(1)}$]} and the intra-layer \txc{[$A_{1s}^{(2)}$]} exciton at $\mathbf{Q}=\mathbf{0}$ for the given layer index $(l_+)$.
			(d) Schematic of the isolated bilayer (sample S1) WS\textsubscript{2} on Si/SiO\textsubscript{2} substrate.
			(e) Temperature dependent ($4.2$ to $295$ K) differential reflectance of bilayer WS\textsubscript{2} \txc{governed by the intra-layer exciton}. The dashed lines show the fitting of the spectra obtained from the model.
			(f) Color plot of $\frac{d}{dE}$$(\frac{\Delta R}{R})$ as a function of energy and temperature.
			(g) The normalized oscillator strength of the \txc{intra-layer} $A_{1s}$ exciton obtained from the fitting as a function of temperature.
}
\label{fig:temp_spectra}
\end{figure}

In the 2H form of bilayer WS\tb2, the individual layers are rotated by {180$^{\circ}$} with respect to each other, making the unit cell inversion symmetric. This requires an additional layer degree of freedom ($l_{\pm}=\pm 1$) to describe excitons in bilayer for a given spin ($s_z=\pm 1$) and valley ($\tau_z=\pm 1$). To describe the lowest energy excitonic states in a bilayer, we use a simple quasiparticle Hamiltonian using tungsten $d$ orbitals as the basis functions \cite{gong2013magnetoelectric,das2019layer} and solve the Bethe-Salpeter (BS) equation\cite{wu2015exciton,das2019layer}, as described in \textbf{Supporting Information 1}. As an example, in Figure \ref{fig:temp_spectra}a, we schematically illustrate the lowest energy conduction and valence bands for one of the degenerate cases ($l_+$) along with the layer distribution of the wave functions close to $\mathbf{K}$ point. The energy dispersion of lowest energy excitonic states with center of mass momentum ($\mathbf{Q}$) is schematically depicted in Figure \ref{fig:temp_spectra}b. The calculated $\mathbf{k}$-space distribution of the excitonic state with $\mathbf{Q}=\mathbf{\mathbf{0}}$ is depicted in Figure \ref{fig:temp_spectra}c. Depending on the layer distribution of the single particle wave functions, we have both intra-layer [$A_{1s}^{(2)}$] and inter-layer [$A_{1s}^{(1)}$] 1s excitons, with the latter being the lowest energy one for bilayer WS\tb2. The layer index of the intra-layer exciton is coupled with spin and valley such that $l_{\pm}$=$s_z$.$\tau_z$=$\pm 1$ \cite{das2019layer}, which results in two degenerate configurations for each at zero external field. While radiative recombination is allowed for both these types of excitons by standard selection rules, the intra-layer exciton exhibits more than an order of magnitude higher radiative decay rate compared with the inter-layer exciton \cite{das2019layer}, and hence the optical response in bilayer \txc{discussed here is dominated by the intra-layer $A_{1s}^{(2)}$} exciton \cite{schuller2013orientation}. The inter-layer $A_{1s}^{(1)}$ exciton noted above is momentum-direct with $\mathbf{Q}=\mathbf{\mathbf{0}}$ \cite{das2019layer,arora2017interlayer,horng2018observation,niehues2019interlayer,gerber2019interlayer}, which is distinct from the indirect inter-layer exciton \cite{wang2018electrical,scuri2020electrically}. Due to the difference in spin splitting of the bright and the dark states in Mo and W-based TMDCs, the inter-layer exciton is energetically above the intra-layer exciton in MoTe\tb2 \cite{arora2017interlayer}, MoSe\tb2 \cite{horng2018observation}, and MoS\tb2 \cite{niehues2019interlayer,gerber2019interlayer,leisgang2020giant}, while it is below the intra-layer exciton in WS\tb2 \cite{das2019layer}.

To extract the spectral feature of the intra-layer exciton in isolated bilayer WS\tb2 transferred on 285 nm SiO\tb2/Si substrate (sample S1 as schematically shown in Figure \ref{fig:temp_spectra}d, see \textbf{Supporting Information 2} for Raman Characterization), we perform differential reflectance measurement (see \textbf{Methods} for measurement details). Figure \ref{fig:temp_spectra}e shows the temperature dependent differential reflectance ($\frac{\Delta R}{R}=\frac{R_{on}-R_{off}}{R_{off}}$) in the range of $4.2$ to $295$ K. Throughout the text, $R_{on}$ denotes the reflected intensity from full stack and $R_{off}$ denotes the same from the stack without the bilayer WS\tb{2}. For notational simplicity, we denote experimentally observed neutral exciton and charged trion peaks as $X^0$ and $X^\pm$, respectively.
The optical response and the line shape are modified by the interference between WS\tb2 and substrate reflection, and the excitonic response of the bilayer is deconvoluted from the overall response. To reproduce the reflectance spectra obtained from the experiment, the exciton contribution to the dielectric response is represented as Lorentzian oscillator \cite{li2014measurement,arora2015excitonic,yu2017giant}:
\beq
\epsilon(E)=\epsilon_{\infty}+\sum_{j}^{}\frac{\lambda_j}{E_j^2+E^2-i\gamma_jE}
\eeq
where the $\epsilon$\tb{$\infty$} is the background response, $\lambda_j$ and $E_j$ are the oscillator strength and the position of the $j^{th}$ oscillator respectively with a broadening parameter $\gamma_j$. $R_{on}$ and $R_{off}$ are then calculated using transfer matrix method (see \textbf{Supporting Information 3} for more details). The color plot of $\frac{d}{dE}$$(\frac{\Delta R}{R})$ in Figure \ref{fig:temp_spectra}f shows a conspicuous spectral red shift of the \txc{neutral $X^0$ exciton} with an increase in temperature, which results from a corresponding reduction in the bandgap. The corresponding temperature tunability  of the exciton oscillator strength is shown in Figure \ref{fig:temp_spectra}g.

\begin{figure}[!hbt]
	\centering
	\vs{-0.1in}
	\includegraphics[scale=0.45]{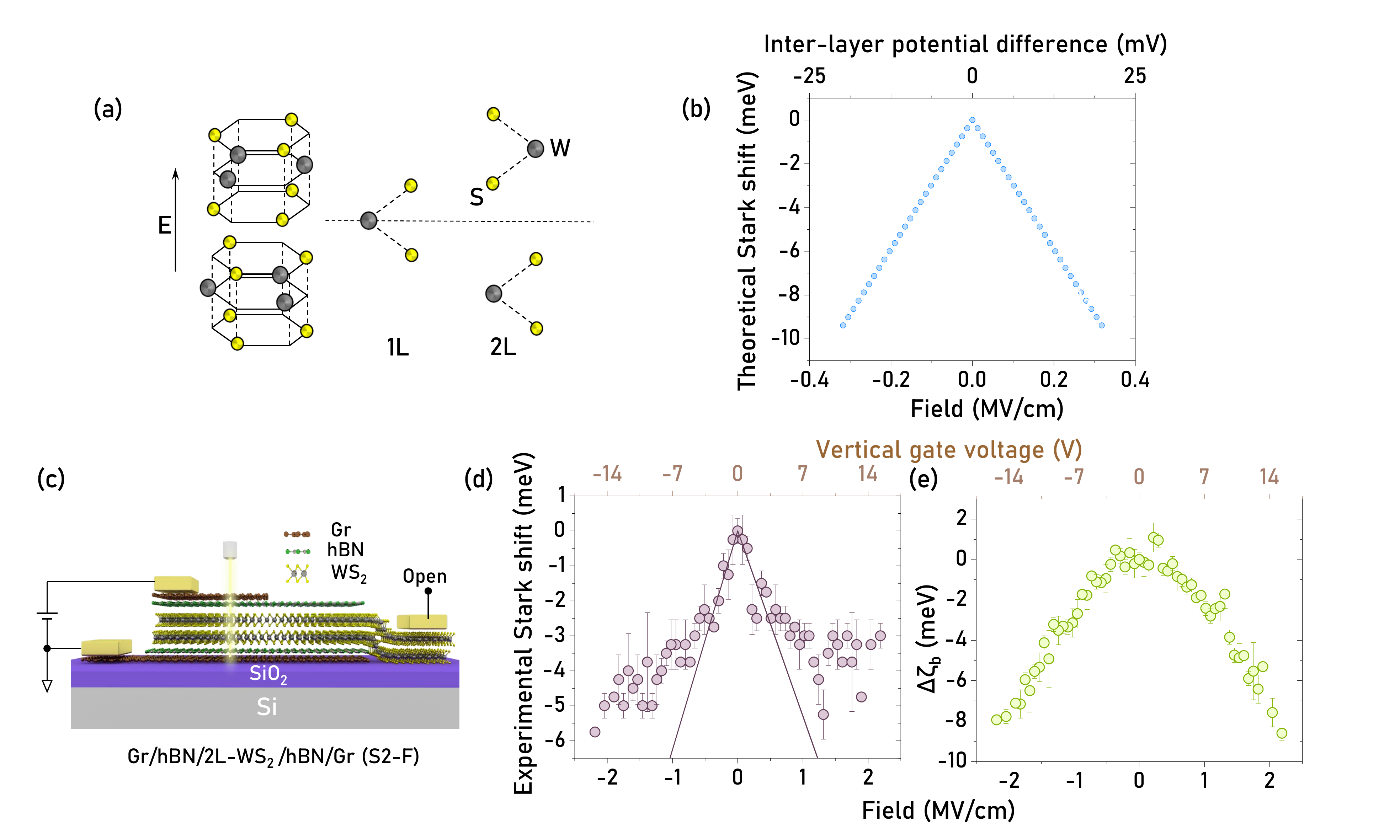}
	\caption{\textbf{Linear QCSE for bilayer WS\tb2.}
		(a) Schematic diagram of the broken reflection symmetry in a bilayer WS\textsubscript{2} along the c-axis while it is maintained for monolayer.
		(b) Calculated electric field dependent linear shift in \txb{intra-layer 1s} exciton position obtained form the BS equation. Here the top axis shows the inter-layer potential difference while the bottom axis denotes the corresponding electric field.
		(c) Schematic diagram of the device sample S2-F where bilayer WS\textsubscript{2} is sandwiched between hBN and graphite layers. The vertical field is applied between the top and the bottom graphene layers while the WS\tb2 layer is kept electrically floating.
		(d) The experimental linear Stark shift (in symbols) for the bilayer device. The solid line represents a linear fit to the Stark shift at low field regime.
		(e) Change in the binding energy ($\Delta \zeta_b$) of the exciton with electric field.
	}

\label{fig:stark_d1}
\end{figure}

Due to the reflection symmetry in a monolayer WS\tb2 (Figure \ref{fig:stark_d1}a), in presence of an out-of-plane electric field, the resulting spatial symmetry of the wave function along the c-axis forces the first order perturbation term of the energy eigenvalue to be zero. This causes a zero linear Stark shift in monolayer, and the second order perturbation provides a quadratic QCSE, as verified experimentally \cite{verzhbitskiy2019suppressed,roch2018quantum,massicotte2018dissociation,chakraborty2019electrically}. However, the reflection symmetry is broken in a bilayer case (Figure  \ref{fig:stark_d1}a) and the resulting asymmetry in the layer distribution of the intra-layer exciton is clearly visible in Figure  \ref{fig:temp_spectra}a-b. This in turn results in a nonzero first order perturbation term, and thus a linear QCSE. The electric field dependent linear shift in the \txb{intra-layer} 1s exciton position as obtained from the solution of the BS equation (see \textbf{Supporting Information 1} for calculation details) is shown in Figure \ref{fig:stark_d1}b. We do not consider any additional screening of the electric field inside the bilayer.

To demonstrate the linear QCSE experimentally, \txc{and to segregate the electric field effect from the doping effect,} we prepare a FLG/hBN/2L-WS\tb2/hBN/FLG (sample S2) stack as schematically shown in Figure \ref{fig:stark_d1}c (see \textbf{Methods} for sample preparation). The FLG layers act as transparent gate electrodes in the back reflection geometry. The thickness of the top and the bottom hBN layers is 20 nm each. The high quality of the hBN layers allows us to apply high electric field without appreciable gate leakage (see \textbf{Supporting Information 4}). \txc{In the field-configuration (denoted as S2-F), a voltage ($V_g$) is applied at the top gate electrode keeping the bottom gate grounded. The electrode directly contacting the bilayer WS\tb2 is kept in open circuit mode, avoiding carrier injection and thus the WS\tb2 is electrically floating. This allows us to study the electric field effect without any confounding electrostatic doping introduced by $V_g$. In this mode of operation, the lack of gate-induced or photo-induced doping \cite{verzhbitskiy2019suppressed} in our sample is evidenced by the absence of any signature of trion in the entire $V_g$ range. The top FLG layer partially covers the device and the sample is illuminated in the FLG-covered portion (Figure \ref{fig:stark_d1}c).} \txc{$\frac{\Delta R}{R}$ and $\frac{d}{dE}$$(\frac{\Delta R}{R})$} for the stack are provided in \textbf{Supporting Information 5} at different electric fields, measured at 4.2 K. The extracted $X^0$ peak shift ($\Delta E_{X^0}$) with respect to zero field is plotted in Figure \ref{fig:stark_d1}d as a function of $V_g$ (top axis) and the vertical electric field $\xi$ (bottom axis). The electric field inside the bilayer WS\tb2 is estimated as:
\beq\label{eq:field}
\xi=\frac{V_g}{d_{w}+d_{h}\frac{\epsilon_{\perp,0,w}}{\epsilon_{\perp,0,h}}}
\eeq
where $d_{w(h)}$ is the thickness of the bilayer WS\tb{2} (hBN), $\epsilon_{\perp,0,w}$ ($=6.3$) and $\epsilon_{\perp,0,h}$ ($=3.76$) \cite {laturia2018dielectric} are the static out-of-plane dielectric constants of WS\tb{2} and hBN, respectively. We find that $\Delta E_{X^0}$ varies linearly with $\abs{\xi}$ - a direct evidence of linear QCSE in bilayer. The average slope of the lines fitting the Stark shift in Figure \ref{fig:stark_d1}d suggests an out of plane static dipole moment value of $\sim 5.8$ meV/MVcm$^{-1}$. The predicted electric field required to provide the same shift as in the experiment is less in the calculation, which indicates a strong screening \cite{santos2013electrically} in the bilayer WS\tb2, particularly in the presence of optical excitation.

At higher $\abs{\xi}$, the experimental Stark shift deviates from linearity as the rate of shift reduces, pointing to strong excitonic effects\txb{\cite{abraham2020anomalous}}  in the QCSE. The exciton binding energy ($\zeta_b$) reduces at higher electric field, which results in a blue shift in the $X^0$ position, partially compensating the QCSE induced red shift. In Figure \ref{fig:stark_d1}e, we plot the difference between the shift predicted from the fitted line and the experimental values  (from Figure \ref{fig:stark_d1}d), which is an estimate of $\Delta \zeta_b$ with an increase in $\abs{\xi}$. The reduction in $\zeta_b$ is attributed to two many-body effects. First, an electric field in the out of plane direction forces the electron and hole to move away from each other \cite{miller1984band,polland1985lifetime}, reducing the spatial overlap of the redistributed wave functions. The second effect arises due to field induced orientation of the excitons in the out of plane direction. The resulting proximity of the particles with similar polarity enhances exciton-exciton repulsion, causing an excitonic blue shift \cite{sie2017observation,unuchek2019valley}.

\begin{figure}[!hbt]
		\centering
		\vs{-0.1in}
		\includegraphics[scale=0.5]{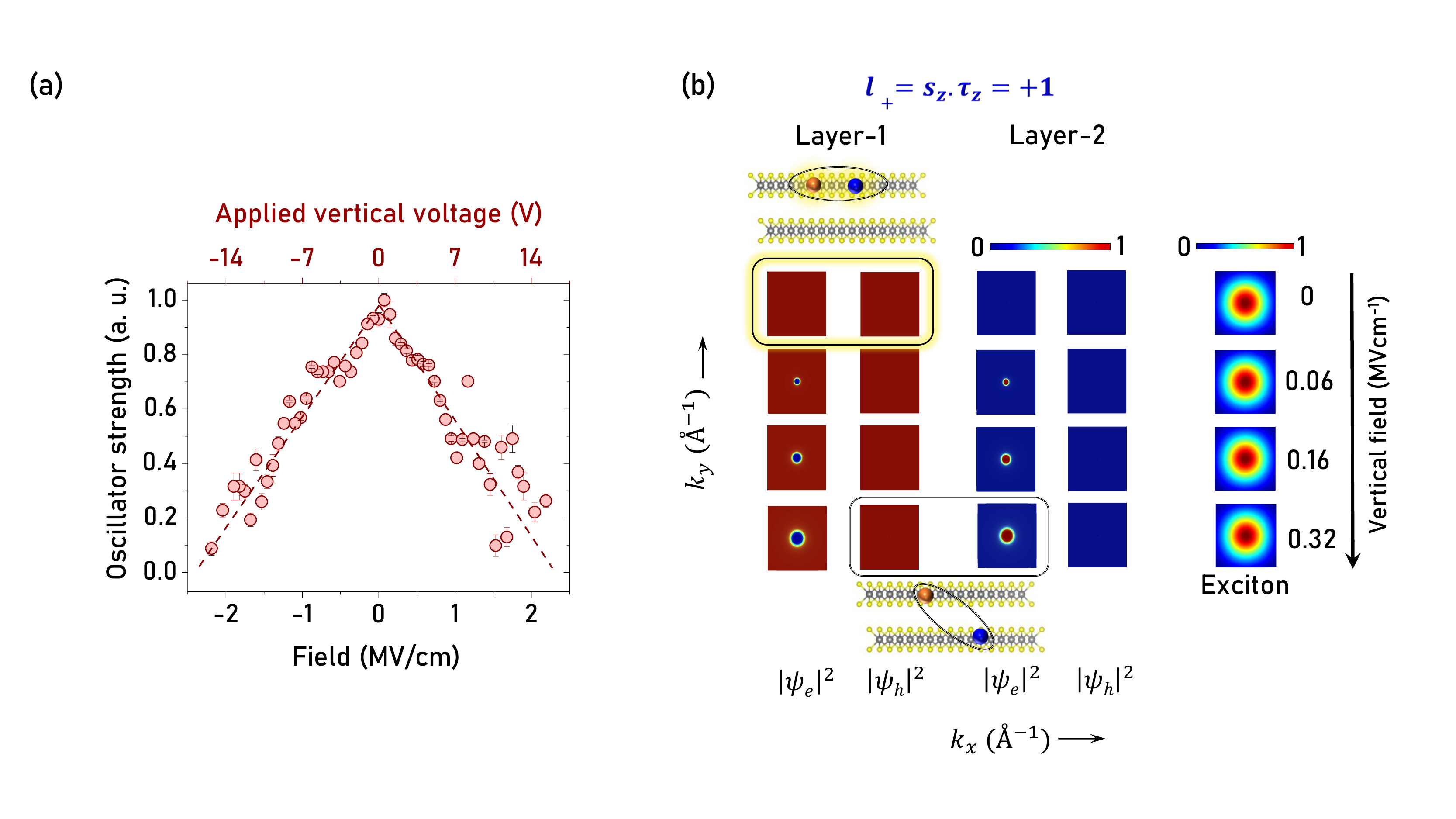}
		\caption{\textbf{Modulation in oscillator strength with electric field due to intra- to inter-layer conversion of excitons.}
			(a) The extracted oscillator strength (normalized) for the sample S2 with the applied vertical electric field. The dashed line shows a guide to eye.
			(b) $\mathbf{k}$ space distribution of the contribution of electron $(\abs{\psi_e(\mathbf{k})}^2)$ and hole $(\abs{\psi_h(\mathbf{k})}^2)$ for a given layer index ($l_+$) at different inter-layer potential (or electric field) values. The corresponding $\mathbf{k}$ space distribution of the exciton at $\mathbf{Q=0}$ is presented in the right panel, indicating largest contribution from regions close to the $\mathbf{K}$ point. As the vertical field increases, the electron wave function moves away from the $\mathbf{K}$ point and the inter-layer character increases.
}
\label{fig:theory}
\end{figure}

Figure \ref{fig:theory}a shows that the extracted oscillator strength ($\lambda$) of $X^0$ sharply reduces with an increase in $\abs{\xi}$, which is promising for atomically thin tunable electro-optic devices. Such a reduction in $\lambda$ with increasing $|\xi|$ is in contrast with monolayer samples, where the photoluminescence intensity is usually a weak function of the vertical field. Since the reduction in binding energy $(\Delta \zeta_b)$ discussed above is quite small compared with $\zeta_b$, the exciton remains vertically confined in the bilayer \cite{miller1984band,polland1985lifetime} and thus dissociation of the exciton under the applied field is an unlikely possibility. As we do not observe any trion peak at higher $\abs{\xi}$, in congruence with the floating WS\tb2 channel, transfer of oscillator strength from neutral exciton to trion is also not plausible. We attribute this reduction of $\lambda$ to a field driven conversion of intra-layer exciton to inter-layer exciton.
The simple Hamiltonian described in \textbf{Supporting Information 1} is again useful to get a qualitative understanding of the underlying physics. \txc{At the left and the middle panels of Figure 3b, we plot, in the two-dimensional $\mathbf{k}$ space centered around $\mathbf{k}=\mathbf{K}$, the fractional contribution of the individual layers to the electron $(\abs{\psi_e(\mathbf{k})}^2)$ and the hole $(\abs{\psi_h(\mathbf{k})}^2)$ constituting the $l_+$ exciton at $\mathbf{Q=0}$ for different vertical fields. The $\mathbf{k}$-space distribution of the exciton in the right most panel suggests that the maximum contribution of the 1s exciton is from the states close to $\mathbf{k}=\mathbf{K}$. At $\xi=0$, the bright exciton arises from electron and hole both lying at the same layer (intra-layer exciton at layer 1 in the top row of Figure 3b). However, with an increase in $\xi$, the contributing electron concentration around $\mathbf{k}=\mathbf{K}$ is pushed to the layer 2, while the hole distribution is maintained at layer 1 (bottom three rows). Since there is a large weight of the electronic states close to $\mathbf{k}=\mathbf{K}$ in the formation of the $\mathbf{Q}=\mathbf{0}$ exciton, this gives rise to an increasing inter-layer character of the exciton. This, in turn, reduces the oscillator strength with an increase in the vertical electric field.}

The intra-layer $A_{1s}^{(2)}$ exciton in bilayer WS\tb2 has fast non-radiative scattering channels to lower lying inter-layer $A_{1s}^{(1)}$ exciton at the $\mathbf{K} (\mathbf{K^\prime})$ point \cite{das2019layer} and to the indirect band edge states. The relaxation channel can be further aided if the bilayer WS\tb2 is stacked on a few layer graphene, due to ultra-fast inter-layer carrier transfer processes \cite{ceballos2014_ACSnano,hong2014_Naturenano,kallatt2019interlayer,jauregui2019electrical,das2020gate} (top panel of Figure 4a). This limits the possibility of re-radiation making the bilayer WS\tb2/graphene stack an excellent atomically thin tunable absorber when excited at the exciton resonance. \txc{The ultra-fast charge transfer from WS\tb2 to graphene also results in a strong de-doping of the WS\tb2 layer.} This is investigated in a stack of FLG/2L-WS\tb{2}/hBN/FLG (sample S3), as schematically illustrated in the bottom panel of Figure 4a. The temperature dependent $\frac{\Delta R}{R}$ and the corresponding oscillator strength at different temperatures are shown in \textbf{Supporting Information 6}. Figure 4b depicts the $\frac{d}{dE}$$(\frac{\Delta R}{R})$ in a color plot at different vertical electric fields. The extracted excitonic red shift and the reduction in the oscillator strength are shown separately in Figures 4c-d. While the linearity in the Stark shift is maintained in this structure as well, we observe about $4$-fold reduction in the slope ($1.48$ meV/MVcm$^{-1}$) compared with \txc{device S2-F}, requiring a higher electric field to obtain a similar Stark shift. This can be correlated with the additional screening in the bilayer WS\tb2 due to the proximity of the graphene layer.

We now investigate the role of electrostatic doping induced by the gate in tuning the \txc{intra-layer} exciton. \txc{For this purpose, we again use sample S2, however, now in a doping configuration (denoted as S2-D). In this mode of operation, the top gate electrode is electrically open circuited, and the metal pad directly contacting the WS\tb2 is electrically grounded. The bottom FLG layer is used as the gate electrode, and a voltage is applied with respect to the grounded electrode, as shown in Figure 5a.} The metal contact can efficiently inject both electrons and holes into the bilayer WS\tb2, depending on the polarity of the gate voltage ($V_g$). Figure 5b shows a two-dimensional color plot of the first derivative of the differential reflectance as a function of $V_g$ (individual spectra are presented in \textbf{Supporting Information 7}). At low $\abs{V_g}$, the optical response is dominated by $X^0$, while at higher positive (negative) $V_g$, $X^-$ ($X^+$) dominates the response, red shifting the resonance energy.

The small blue shift ($\approx 2-3$ meV) of the exciton with an increase in $\abs{V_g}$ (Figure 5c) can be attributed to a combined effect of (a) renormalization of the excitonic gap due to the increased doping \cite{chernikov2015electrical,yao2017optically}, (b) reduction in exciton binding energy due to separation of electron-hole wave function along the c-axis by the gate field \cite{miller1984band,polland1985lifetime}, and (c) reduction in the exciton binding energy due to enhanced exciton-exciton repulsion due to oriented exciton dipoles \cite{sie2017observation,unuchek2019valley}. On the other hand, the trion position red shifts with an increase in $\abs{V_g}$, leading to an increased separation between the exciton and trion resonance energies (Figure 5d). Similar observations are reported in monolayer as well \cite{xu2014spin,li2019emerging}, which is usually attributed to the enhancement in the energy required to move the additional electron (for $X^-$) or hole (for $X^+$) to respective bands due to doping induced Pauli blocking \cite{mak2013tightly,kallatt2019interlayer}.

The extracted oscillator strength of $X^0$ and $X^\pm$ in Figure 5e shows a doping dependent transfer of oscillator strength from one excitonic species to another. At low $V_g$, when the bilayer Fermi level is deep inside the band gap, $X^0$ formation is favoured. Efficient carrier injection in bilayer allows us to sweep the Fermi level almost the entire bandgap, providing strong n-type (p-type) doping at positive (negative) $V_g$, which favours the formation of $X^-$ ($X^+$), suppressing the formation of $X^0$. Such a near complete modulation of oscillator strength of different excitons is promising for electrically tunable ultra-thin optoelectronic components.

We observe a sharp increase in the extracted oscillator strength for both $X^0$ and $X^\pm$ with an increase in $\abs{V_g}$ (Figure 5f). For $X^0$, this is attributed to an enhanced Coulomb scattering in the presence of large electron or hole density, and neutral exciton to trion conversion. The charged nature of the trion species makes the Coulomb scattering stronger resulting in a larger broadening. Similar enhancement in broadening has been observed previously in III-V quantum wells under vertical electric field \cite{miller1984band}, and has been attributed to increased rate of dissociation of exciton in presence of the electric field. However, in the present case, both the exciton and trion binding energy are significantly larger than their III-V quantum well counterpart. In addition, a stronger quantum confinement in the out of plane direction forces the change in the binding energy to be relatively small, as discussed previously. It thus appears that dissociation induced increase in broadening may play a negligible role in our experiment, particularly for the neutral exciton.

In summary, We propose bilayer WS\tb2 as an atomically thin, tunable electro-absorption layer when excited at the excitonic resonance. The spectral feature as well as the oscillator strength of the excitonic response is highly tunable by an external gate voltage through both electrostatic doping and quantum confined Stark effect. Contrary to the monolayer case, the quantum confined Stark effect in bilayer shows a linear variation with electric field owing to reflection symmetry breaking, coupled with a large modulation in the oscillator strength due to \txb{a partial} inter-conversion between intra-layer and inter-layer exciton. The electrically tunable nature of the excitonic response has interesting prospects for ultra-thin, light and compact, fast, reconfigurable  optoelectronic applications like modulators, pulse shapers, photodetectors, digital mirrors and phased arrays, tuning opacity of surfaces, and possible elements for optical and quantum information processing.
\section*{Methods}
\textbf{Device fabrication:} The heterojunctions are prepared using dry transfer of individual exfoliated layers to a Si substrate coated with 285 nm SiO\tb2 layer. The transfer of each layer is performed one at a time under a microscope with controlled translational and rotational stages for precise alignment. After transfer of each layer, the sample is heated on a hot plate at $80^\circ$ C for 2 minutes for improved adhesion between layers. Optical images of samples S2 and S3 are shown in \textbf{Supporting Information 8} after all the layers are transferred. Standard nanofabrication techniques are used to define the metal electrodes contacting the graphene layers. The substrate with the transferred layers is spin coated with PMMA 950C3 and baked on a hot plate at $180^\circ$ C for 2 minutes. This is followed by electron beam lithography with an acceleration voltage of 20 KV, an electron beam current of 210 pA, and an electron beam dose of 220 $\mu$Ccm\tu{-2}. Patterns are developed using MIBK:IPA solution in the ratio 1:3. Later samples are washed in IPA and dried in N\tb{2} blow. Electrodes are made with 10 nm Ni /50 nm Au films  deposited by DC magnetron sputtering at $3\times10^{-3}$ mBar. Metal lift-off is done by dipping the substrate in acetone for 20 minutes, followed by washing in IPA and N\tb{2} drying.

\textbf{Reflectance measurement:} For temperature dependent micro-reflectance measurements, the samples are placed in a closed cycle He cryostat with an optical window placed above the sample. The samples are illuminated with a broadband white LED source and response is collected through a $\times 50$ objective with a numerical aperture of 0.5 and analyzed using a spectrometer with a grating of 1800 lines/mm. The differential reflectance is then calculated via normalizing the signal obtained from stack with ($R_{on}$) and without bilayer WS\tb2 ($R_{off}$) layer as follows: $\frac{\Delta R}{R}=\frac{R_{on}-R_{off}}{R_{off}}$. To match the experimental data, apart from the $A_{1s}$ exciton, we consider oscillators separated by $50$ meV in the spectral range below the exciton while the higher energy side is fitted with oscillators placed at  $A_{2s}$, $A_{3s}$, $A_{4s}$, $A_{5s}$, and  $B_{1s}$ energy positions \cite{he2014tightly,chernikov2014exciton}.

\section*{Supporting Information}
S1. Stark shift calculation of the excitons in bilayer TMD, S2. Raman characterization of bilayer WS\tb2, S3. Calculation of reflectance with transfer matrix method, S4. Gate leakage current for sample S2, S5. Field dependent differential reflectance for the device configuration S2-F, S6. Temperature dependent reflectance spectra for sample S3, S7. Gate dependent reflectance spectra for device configuration S2-D, S8. Optical images of the stacks.
\section*{Acknowledgements}
K. M. thanks Varun Raghunathan for useful discussions. K. M. acknowledges the support a grant from Indian Space Research Organization (ISRO), a grant from MHRD under STARS, grants under Ramanujan Fellowship and Nano Mission from the Department of Science and Technology (DST), Government of India, and support from MHRD, MeitY and DST Nano Mission through NNetRA. K.W. and T.T. acknowledge support from the Elemental Strategy Initiative conducted by the MEXT, Japan, Grant Number JPMXP0112101001, JSPS KAKENHI Grant Numbers JP20H00354 and the CREST(JPMJCR15F3), JST.
\section*{Competing Interests}
The Authors declare no Competing Financial or Non-Financial Interests.

	\bibliography{references}
	\newpage

\newpage

\newpage

\newpage
\begin{figure}[!hbt]
	\centering
	\vs{-0.1in}
	\includegraphics[scale=0.5]{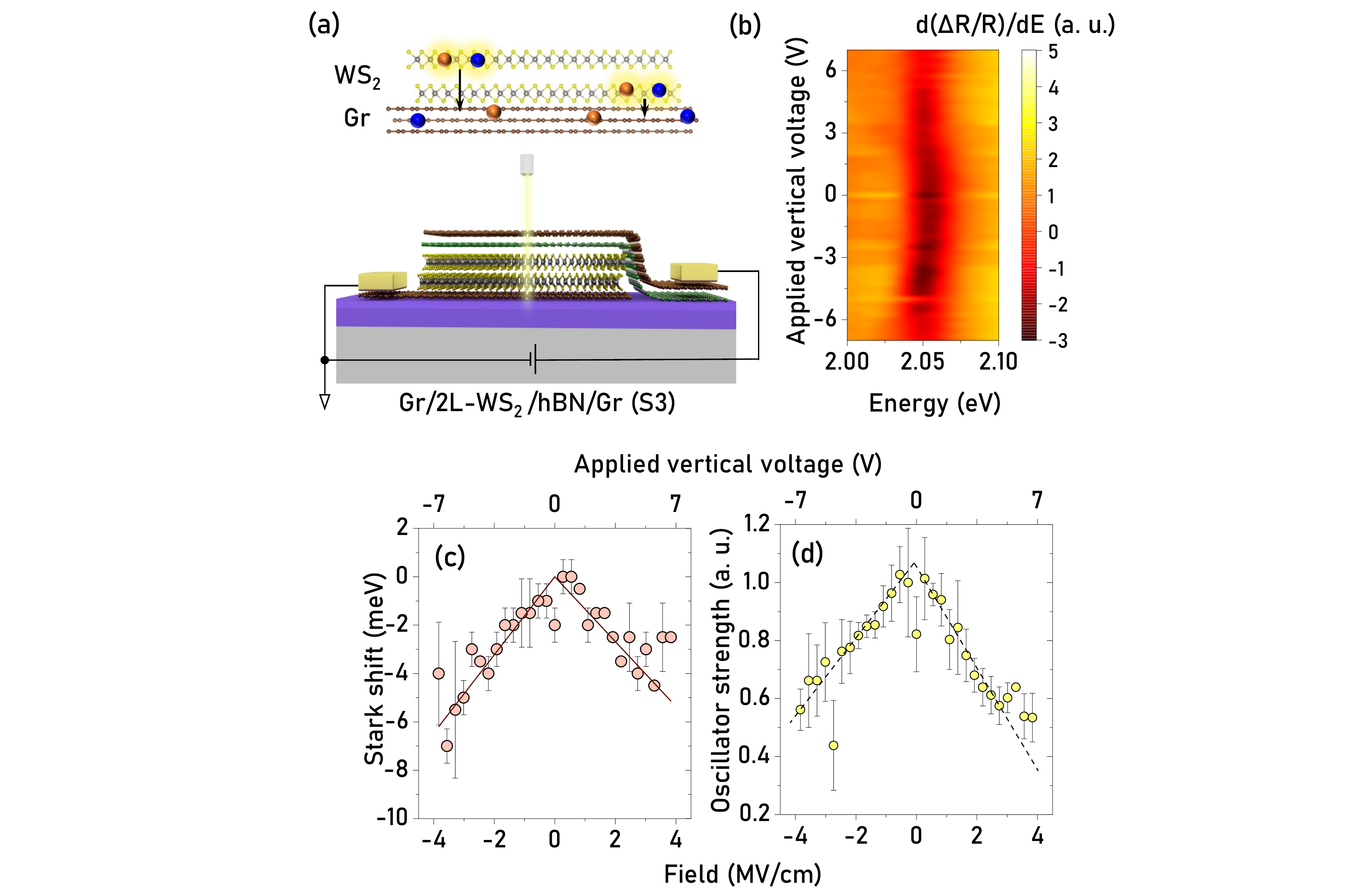}
	\caption{\textbf{QCSE of bilayer excitons directly placed on few-layer graphene.}
		(a) Top panel: A cartoon picture of the ultra-fast inter-layer transfer process from WS\tb2 to few-layer graphene. Bottom panel: Schematic diagram of the device (sample S3) where bilayer WS\tb2 is directly placed on graphene. Bias is applied between the two graphene layers.
		(b) 2D Color plot of $\frac{d}{dE}$$(\frac{\Delta R}{R})$ from sample S3 across the applied voltages. The exciton resonance is red shifted at higher field and the corresponding oscillator strength decreases.
		(c-d) The Stark shift and the normalized oscillator strength of WS\tb2 as a function of the applied vertical field (bottom axis). The top axis shows the applied voltage. The solid lines in (c) represent a linear fit to the data, and the dashed lines in (d) are guides to eye.
			}
	
	\label{fig:stark_effect_d2}
\end{figure}

\newpage
\begin{figure}[!hbt]
		\centering
		\vs{-0.1in}
		\includegraphics[scale=0.5]{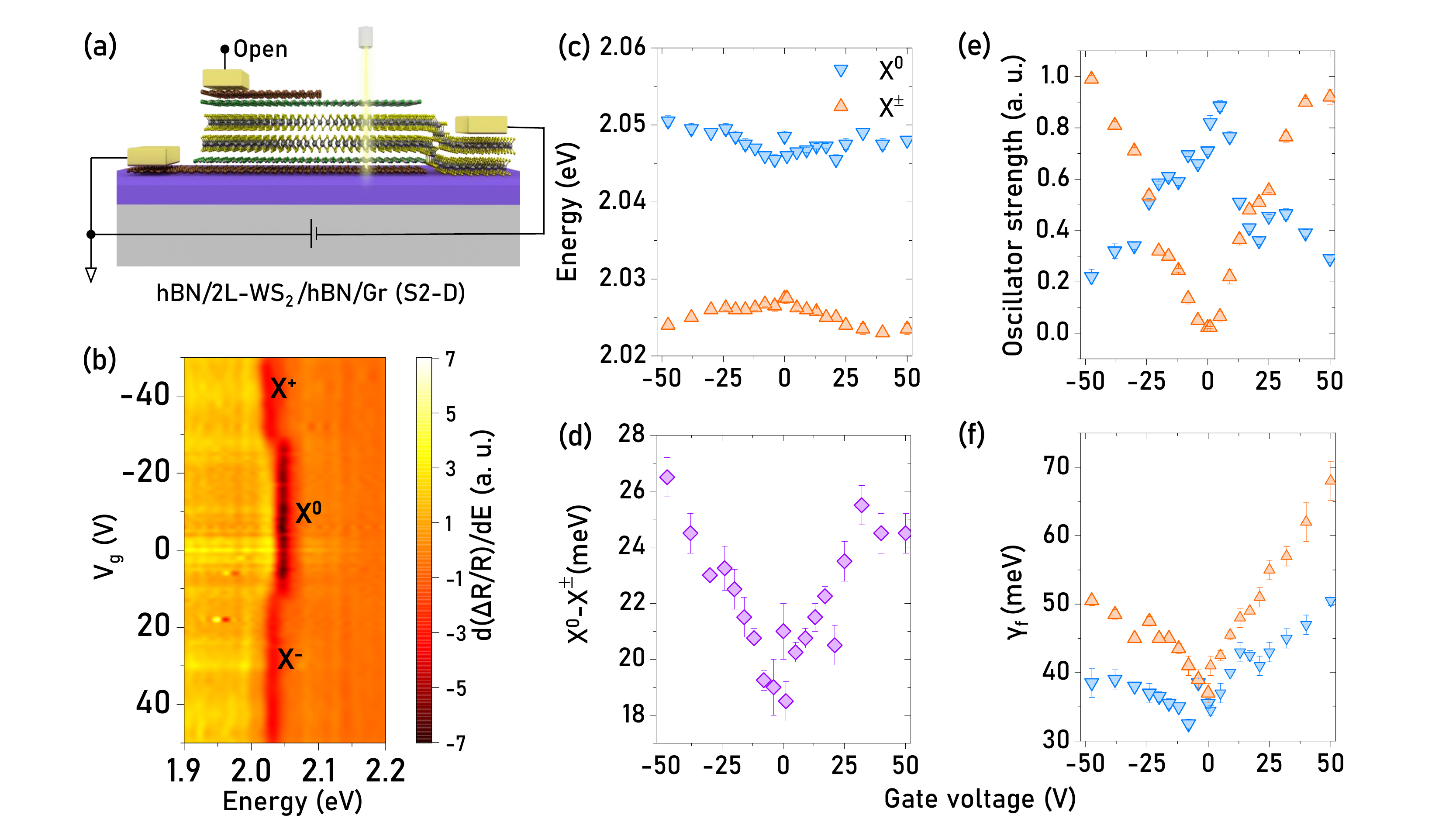}
		\caption{\textbf{Electrostatic doping induced tuning of excitons and trions in bilayer WS\tb2.}
			(a) Schematic of the device \txc{(S2-D)} used for gating analysis. Here the bilayer WS\textsubscript{2} is sandwiched between hBN layers but directly connected to one of the electrode. Voltage is on the bottom graphene layer keeping the electrode connecting the WS\textsubscript{2} layer grounded.
			(b) 2D color plot of $\frac{d}{dE}$$(\frac{\Delta R}{R})$ as a function of the gate voltage ($V_g$). The exciton ($X^0$) and trion ($X^{\pm}$) dependence with the $V_g$ is also clearly observed.
			(c) $V_g$ dependent $X^0$ and $X^{\pm}$ positions.
			(d) Increase in separation between the $X^0$ and the $X^{\pm}$ with increasing $V_g$.
			(e) The extracted oscillator strength (normalized) of the $X^0$ and $X^{\pm}$ as a function of $V_g$.
			(f) The extracted broadening parameter ($\gamma_f$) with $V_g$ variation.			
	}

		\label{fig:gating}
\end{figure}


\section*{Supplementary material}

# S1. Stark shift calculation of the excitons in bilayer TMD:

The bilayer Hamiltonian (H) with an inter-layer potential difference of $2U$ for $WS_2$ can be written as[1,2]

$$H = \begin{bmatrix} \Delta - U & at_i(\nu_z k_x + ik_y) & 0 & 0 \\ at_i(\nu_z k_x - ik_y) & -\nu_z s_z \lambda - U & 0 & t_\perp \\ 0 & 0 & \Delta + U & at_i(\nu_z k_x - ik_y) \\ 0 & t_\perp & at_i(\nu_z k_x + ik_y) & \nu_z s_z \lambda + U \end{bmatrix} \quad (1)$$

where $\Delta$ is the bandgap, $a$ is the lattice constant, $t_i$ is the nearest-neighbour intralayer hopping, $\lambda$ is the spin-valley coupling for holes in monolayer, $t_\perp$ is the interlayer hopping for holes, $\nu_z$ is the valley degree of freedom, and $s_z$ is spin degree of freedom ($\pm 1$). Material parameters are obtained from ref.[1]

An exciton state $(|\Psi\rangle_{\vec{Q}})$ in an exciton band $n$ with center of mass momentum $\vec{Q}$ in reciprocal space can be written as[3]

$$|\Psi\rangle_{\vec{Q}} = \sum_{v,c,\vec{k}} A^n_{v,c,\vec{Q}}(\vec{k}) \left|v,\vec{k}\right\rangle \left|c,\vec{k}+\vec{Q}\right\rangle \quad (2)$$

$A^n_{v,c,\vec{Q}}(\vec{k})$ can be obtained from the solution of the Bethe-Salpeter (BS) equation[3]

$$\left\langle v,c,\vec{k},\vec{Q}\right| H \left|v',c',\vec{k}',\vec{Q}\right\rangle = \delta_{vv'}\delta_{cc'}\delta_{\vec{k}\vec{k}'}(\varepsilon_{(\vec{k}+\vec{Q})c} - \varepsilon_{\vec{k}v}) - (\Xi - \Lambda)^{cc'}_{vv'}(\vec{k},\vec{k}',\vec{Q}), \quad (3)$$

Here $\varepsilon$ is the eigenvalue obtained by diagonalizing the quasiparticle Hamiltonian described above. $\Xi$ and $\Lambda$ are the direct and exchange term in the two-particle matrix elements. We take $\Lambda = 0$ for excitons with $Q = 0$.

To obtain the Stark effect, the above equation is solved for different $U$, and the corre-

sponding exciton energy eigenvalues at $Q = 0$ are calculated. The corresponding vertical field ($\xi$) is calculated as

$$\xi = 2U/t_0, \qquad (4)$$

where $t_0$ is the physical separation between two Tungsten layers (6.3Å).

## S2. Raman characterization of bilayer WS$_2$

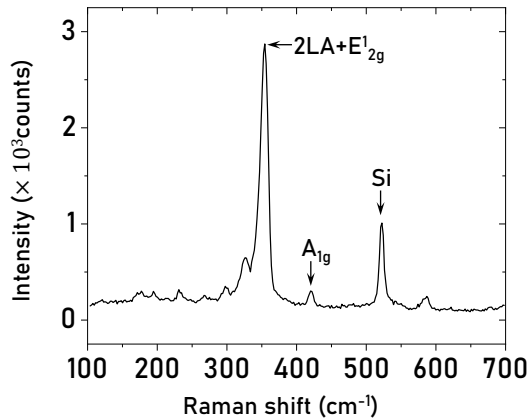

Figure S1: **Raman characterization of bilayer WS$_2$.** Room temperature Raman spectra for bilayer WS$_2$ (S1). Different Raman modes are marked in the figure.

## S3. Calculation of reflectance with transfer matrix method:

The reflectance of the stacks is obtained by calculating the reflection from multiple layer dielectric formalism. For an N layer system the incident electric field ($\mathcal{E}$) can be written as[4]

$$\begin{bmatrix} \mathcal{E}_1^+ \\ \mathcal{E}_1^- \end{bmatrix} = \mathcal{R} \begin{bmatrix} \mathcal{E}_{(N+2)}^+ \\ \mathcal{E}_{(N+2)}^- \end{bmatrix} \quad (5)$$

where $\mathcal{E}_j^+$, $\mathcal{E}_j^-$ are the incoming and outgoing electric field from the $j^{th}$ surface

$$\mathcal{R} = \mathcal{R}_1 \mathcal{R}_2 ... \mathcal{R}_{N+1} \quad (6)$$

and, under normal incidence,

$$\mathcal{R}_j = \frac{1}{\tau_j} \begin{bmatrix} \exp(i\delta_j) & \rho_j \exp(i\delta_j) \\ \rho_j \exp(-i\delta_j) & \exp(-i\delta_j) \end{bmatrix} \quad (7)$$

$$\delta_1 = 0, \delta_j = k_j t_j = (2\pi/\lambda)\tilde{n}_j t_j \quad (8)$$

$$\tau_j = \frac{2\tilde{n}_j}{\tilde{n}_j + \tilde{n}_{j+1}}, \rho_j = \frac{\tilde{n}_j - \tilde{n}_{j+1}}{\tilde{n}_j + \tilde{n}_{j+1}} \quad (9)$$

Here $\tau$ is the transmission coefficient, $\rho$ is the reflection coeffecient and $\tilde{n}$ is the complex refractive index of the individual layers with thickness $t$. Further $\tilde{n}$ defined as $\tilde{n}=n-i\kappa$ with $\kappa$ is the extinction coefficient of the material.

The reflection from the stack is defined by $R = \left|\mathcal{E}_1^-/\mathcal{E}_1^+\right|^2$, which is calculated for the stacks with ($R_{on}$) and without ($R_{off}$) the bilayer WS$_2$ layer. Complex refractive index ($\tilde{n}$) of graphite, Si and SiO$_2$ with wavelength dispersion are taken from literature,[5,6] $\kappa$ of hBN is assumed to be zero with $n$ of 1.85.[7] The $\tilde{n}$ value for WS$_2$ is obtained from the Lorentzian oscillator model described in the main text. To fit the differential reflectance spectra ($\frac{\Delta R}{R}$) obtained from the experiment, we further calculate $\frac{\Delta R}{R} = \frac{R_{on}-R_{off}}{R_{off}}$.

# S4. Gate leakage current

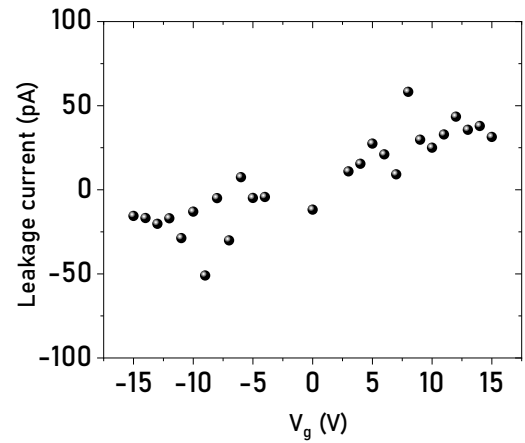

Figure S2: **Gate leakage current for sample S2.** Measured gate leakage current for the experimental range of applied gate voltage for the sample S2.

# S5. Field dependent reflectance spectra for device configuration S2-F

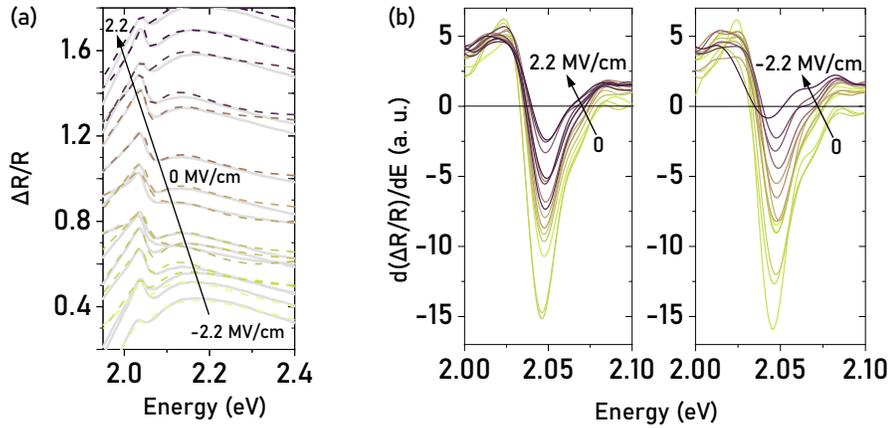

Figure S3: **Field dependent reflectance spectra for device configuration S2-F.** (a) Field dependent differential reflectance for the device configuration S2-F. (b) First derivative of the differential reflectance of bilayer $WS_2$ with the applied vertical field (in MVcm$^{-1}$), measured at 4.2 K. The left and the right panels show the positive and negative electric field, respectively.

# S6. Temperature dependent reflectance spectra for sample S3

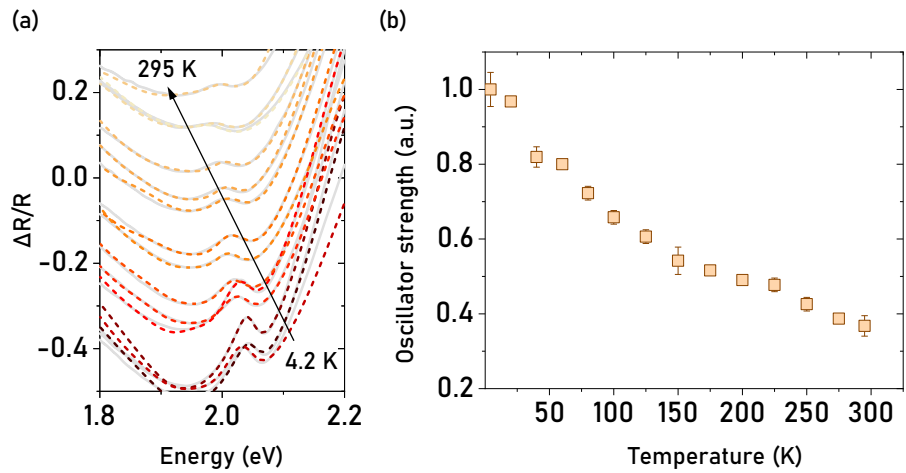

Figure S4: **Temperature dependent reflectance spectra for sample S3.** (a) Temperature dependent differential reflectance for the sample S3. Here the bilayer $WS_2$ is directly sitting on few layer graphene enhancing the fast nonradiative processes. (b) Normalized oscillator strength of the exciton extracted from the fitting across the temperature range.

# S7. Gate dependent reflectance spectra for device configuration S2-D

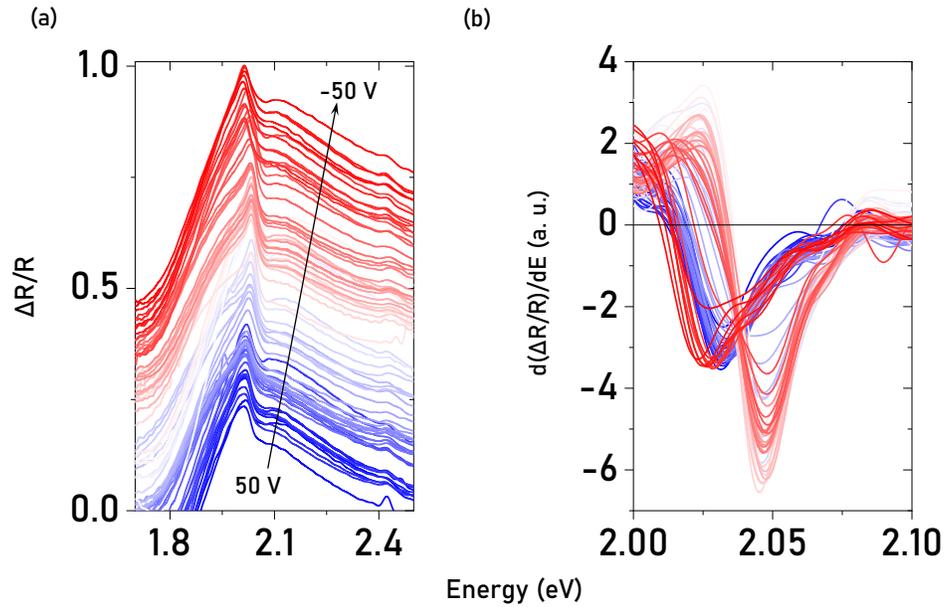

Figure S5: **Gate dependent reflectance spectra for device configuration S2-D.** (a) Gate dependent differential reflectance spectra for the device configuration S2-D showing the electrostatic doping and the transfer of oscillator strength from exciton to trion at higher gate voltage. (b) First derivative of the reflectance spectra in the same voltage range.

# S8. Optical images of the stacks

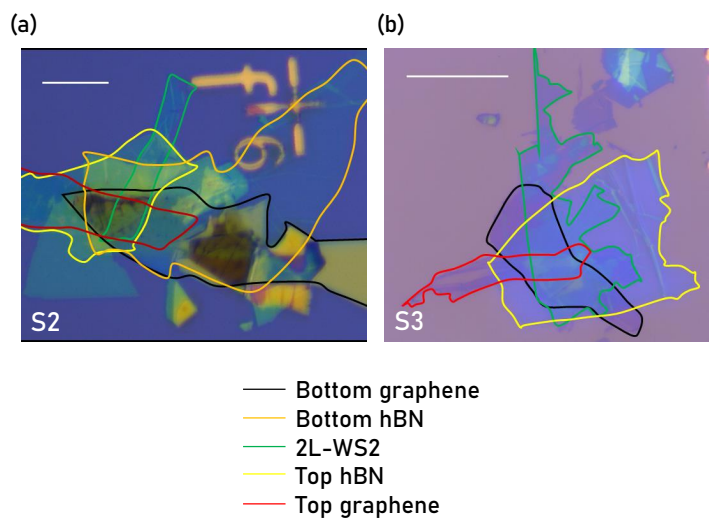

Figure S6: **Optical images of the stacks in sample S2 and S3.** (a-b) Optical image of the complete stacks of sample S2 and S3. The different layer boundaries are marked with different colors. The scale bar is 20 $\mu$m.

	
\end{document}